\documentclass[structabstract]{aa}
\usepackage{float}
\usepackage{lineno}
\usepackage{natbib}
\bibpunct{(}{)}{;}{a}{}{,}
\usepackage{subfigure}
\usepackage{graphicx}
\usepackage{epsfig}
\usepackage{amsmath}
\usepackage[varg]{txfonts}
\usepackage{lscape}
\usepackage{longtable}
\usepackage{txfonts}
\usepackage{color}

\makeatletter

\newcommand{\Rmnum}[1]{\expandafter\@slowromancap\romannumeral #1@}
\makeatother
\begin{document}
   \title{Spectro-polarimetry of the bright side of Saturn's moon Iapetus\thanks{Based on observations made
   with ESO Telescope (UT1) at the Paranal observatory under programme ID 383.C-0058(A), 384.C-0040(A), 385.C-0052(A), and 386.C-0075(A)}}
   \author{C. Ejeta\inst{1, 2}, H. Boehnhardt\inst{1}, S. Bagnulo\inst{3}
           \and G.P. Tozzi\inst{4}
           }
\institute{Max-Planck-Institut f\"ur Sonnensystemforschung,
              Max-Planck-Strasse 2, 37191 Katlenburg-Lindau, Germany\\
              \email{ejeta@mps.mpg.de}\\
 \and  Institut f\"ur Geophysik und extratrrestishe Physik,
        TU Braunschweig, Germany\\
\and Armagh Observatory, College Hill, Armagh BT61 9DG, Northern Ireland, UK\\
\and INAF - Oss. Astrofisico di Arcetri, Largo E. Fermi 5, I-50125 Firenze, Italy}
  \abstract
     {Measurements of the polarized reflected sunlight from atmosphereless solar
     system bodies, over a range of phase angles, provide information about the surface structure and composition.}
   {With this work, we provide analysis of the polarimetric observations of the bright side of Iapetus at five different phase angles, and over
   the full useful wavelength range (400-800nm), so as to assess the light scattering behaviour of a typical surface water
   ice.}
   {Using FORS2 of the ESO VLT, we have performed linear spectro-polarimetric observations of
   Iapetus' bright side from 2009 to 2011 at five different phase angles, in the range from
      $0.80-5.20^{\circ}$, along with circular spectro-polarimetric observations at one phase angle.}
   { By measuring, with high accuracy ($\sim 0.1$\,\% per spectral bin for each Stokes parameter), the spectral polarization of the bright trailing hemisphere of Saturn's
    moon Iapetus, we have identified the polarimetric characteristics of water ice, and found that its linear degree of
    negative  polarization decreases with increasing phase angle of observation (varying from $-0.9$\,\% to $-0.3$\,\%), with a clear dependence on wavelengths of observation.}
   {}
\titlerunning{Spectro-polarimetry of the bright side of Saturn's moon Iapetus}
\authorrunning {Ejeta \textit{et al}.}
   \maketitle
\section{Introduction}
Iapetus has a radius of 718 km and orbits Saturn every 79.33 earth days at a
distance of 3,560,840 km; it has an average bulk density of
1.150$\pm$0.004 g/cm$^{3}$ (http://ssd.jpl.nasa.gov) that might imply that it is made mostly of
ices.
 Tidal interactions with Saturn have synchronized the rotation of Iapetus with its orbital period, and as a result, the moon always
keeps the same face to Saturn and always leads with the same face in its orbital
motion. Iapetus has a striking property that its leading and trailing hemispheres represent a
contrast in surface albedo amounting to a factor of $\sim$ 10
\citep{Squyres:1983}. The leading hemisphere is much darker than its trailing
hemisphere, with an albedo of 0.04 and 0.39, respectively \citep{Spencer10}. There are several hypothesis as to how this global
albedo dichotomy has originated. \citet{Spencer10} have developed the idea put forward pre-cassini
mission by Mendis \& Auxford (1974), suggesting that exogenic
deposition of dark material, possibly from  Saturn's outer retrograde
satellites, has resulted in darkening of the leading hemisphere thereby raising its
temperature and water ice sublimation rates.

\section{Observations and data reduction}

Linear spectro-polarimetric observations of the bright side of Iapetus have been obtained at five different
epochs, using the FORS2 instrument of the European Southern Observatory (ESO) Very Large Telescope (VLT), from April 2009 to March 2011.
FORS is the visual and near UV FOcal Reducer and low dispersion Spectrograph of the ESO VLT  \citep{Appenzeller98}. FORS2, installed on Unit Telescope 1 (Antu), is the version of
FORS in operation since April 2009, and it offers many observing modes, including imaging polarimetry (IPOL) and multi-object spectro-polarimetry (PMOS). In addition to the linear polarization, we have also
performed circular polarization observations of the same side of Iapetus at one phase angle.
  Our observing cadence of Iapetus' bright side polarimetry followed from
the scientific requirement to measure {\textit{separately} its bright hemisphere (to avoid contamination from the dark side), which
is only possible close to its western elongation during its orbit
around Saturn. Hence, all our observations were performed according to the observing blocks prepared
considering a 12 days wide window centered at the date of maximum elongation,
when a rather exclusive viewing geometry of the bright hemisphere of the moon
is guaranteed. During the entire observation epochs, we used the same instrument set up (grism 300V, filter GC435, and a 1$''$ slit width),
with a $\lambda/2$ retarder waveplate (for linear polarization) or $\lambda/4$ (for circular polarization) and a Wollaston prism inserted in the FORS2 optical path
. The Wollaston splits up
 the incoming beam into two rays that are characterized by orthogonal polarization
 states with respect to the orientation (principal plane)
of the Wollaston. The two signals from the Wollaston beams do not overlap as a special aperture masks are
employed for this purpose.

Each epoch of linear polarimetric observations of Iapetus' bright hemisphere were performed  at all position angles of the $\lambda/2$ retarder
waveplate (with respect to the principal plane of the Wollaston prism) from $0^{\circ}$ to $337.5^{\circ}$, in $22.5^{\circ}$ steps; and
at $\lambda/4$ retarder wave plate position angles of 315$^{\circ}$, 45$^{\circ}$,135$^{\circ}$, 225$^{\circ}$ for the circular polarization.
FORS instrument is proved to be affected by a cross-talk problem from linear to circular
polarization \citep{Bagnulo09}. To strongly reduce this linear to circular cross talk, we measured a unique
circular polarization signal as an average of the values obtained at instrument position angles that differ by 90$^{\circ}$, as suggested by \citet{Bagnulo09}, to
which we refer for further technical details.
For the data reduction, we mainly used ESO FORS pipeline tool \citep[see][]{Izzo10}
but independent manual reduction with MIDAS and IRAF
 packages were also used as a cross-check. We applied bias-subtraction to the
 frames using a master bias obtained from a series of five frames taken closer in time to the observations, then divided
by a flat-field obtained by combining five lamp frames.
In fact, by performing the data reduction with and without flat-fielding, we have verified that the flat fielding
does not change the value of the Stokes profiles, as we measure polarization from the normalized flux difference. For each frame observed at various retarder wave plate positions
$\alpha$, we rebinned spectral points to 10 pixels bin, a value we considered as the best compromise between spectral resolution
loss and increase of signal to  noise ratio, and then we extracted the parallel and
the perpendicular beams ($f^{\parallel}$ and $f^{\perp}$, respectively). The extraction radius we used is 6 pixels (rebinned); but we have also
performed the extraction for larger radii ( 8 pixels and 10 pixels) and
confirmed that the results are practically the same. We then performed independent wavelength
calibration for the two beams of each of the frames using an arc lamp frame reduced in the same way as the science frames.

\subsection{Method of performing spectro-polarimetry}
We measured the reduced Stokes $Q$ , $U$, and $V$ parameters (normalized by intensity) by combining fluxes of the pararallel and the
perpendicular beams, employing the technique called the {\textit{difference method}, implemented in \citet{Bagnulo09}, as:
\begin{equation}
P_{X}=\frac{1}{2N}\sum_{j=1}^{N}\left [ \left ( \frac{f^{\parallel}-f^{\perp}}{f^{\parallel}+f^{\perp}}\right )_{\alpha_{j}}- \left (\frac{f^{\parallel}-f^{\perp}}{f^{\parallel}+f^{\perp}}\right )_{\alpha_{j}+\Delta P}\right]
\end{equation}, where N is the number of pairs of exposures for each Stokes parameters (in our case, 4 for the linear polarization, and 2 for the circular
polarizatiin), and\\
           $-$ for $X$ = $Q$, $\Delta$P = 45$^{\circ}$, and $\alpha_{j}$ belongs to the set \{ $0^{\circ}$, $90^{\circ}$, $180^{\circ}$, $270^{\circ}$\} \\
           $-$ for $X$ = $U$, $\Delta$P = 45$^{\circ}$, and $\alpha_{j}$ belongs to the set \{ $22.5^{\circ}$, $112.5^{\circ}$, $202.5^{\circ}$, $292.5^{\circ}$\}\\
           $-$ for $X$ = $V$, $\Delta$P = 90$^{\circ}$, and $\alpha_{j}$ belongs to the set \{$-45^{\circ}$,
           135$^{\circ}$\},\\
           where $Q/I = P_{Q}$, $U/I = P_{U}$, and $V/I=P_{V}$.

Assuming the flux and the standard deviation in each beam to be approximately equal to $f$ and $\sigma$, respectively \citep[see e.g.,][]{Bagnulo09}
the statistical error can be given as:
\begin{equation}
\sigma_{P_{X}}= \frac{1}{2\sqrt{N}}\frac{\sigma}{f}
\end{equation}

\subsection{Analysis of standard stars for linear polarization}
As part of calibration of our measurements, we have carefully  analysed spectro-polarimetric observations of standard stars
for linear polarization (both the non-polarized ones and those with well known higher polarization values), observed with the same instrument setup, and the same night as our target.
 Figures~\ref{fig7} and \ref{fig8} show the measured reduced Stokes
$P_{Q}$ and $P_{U}$ profiles of two non-polarized standard stars observed during the same
night as our target. From Fig.~\ref{fig7}, it is evident that there is a
polarization value of $\sim 0.10$\,\% both in $P_{Q}$ and $P_{U}$ for the standard star ${\rm HD\,97689}$ (observed during the night 2010-05-04).
This value, obtained after applying correction for the
instrument offset during the respective epoch of the observation (discussed below), accounts for an instrumental contribution to the observed polarization in $P_{Q}$.
Figure~\ref{fig8} shows the reduced Stokes $P_{Q}$ and $P_{U}$ profiles of
the standard star ${\rm WD\,1620391}$ observed during the nights 2010-07-25 and
2011-03-28. In the upper panels (corresponding to the epoch 2010-07-25), it can be seen that there is a polarization value of $\sim
0.10$\,\% both in $P_{Q}$ and $P_{U}$, while in the lower panels (for the epoch 2011-03-28) the respective values are a
bit lower ($\sim0.05$\,\%).
\\From the observations of the polarized standard stars, observed during the same night
as our target, we discovered also that the measured position angles of these stars during some of our observation epochs are different from their corresponding expected
values measured using FORS1 by \citet{Fossati07}. This observed offsets clearly demonstrate that the polarimetric optics of the instrument
 were not correctly aligned during some of our observing epochs.
Hence, we have calculated the FORS2 instrument offsets during some epochs of
our observation, and the values are depicted in Table~\ref{table1}. Figure \ref{fig6} shows such a particular case of our spectro-polarimetric
measurements of the standard star for linear polarization CD-28 13479 (Hiltner
652) that was obtained during the night 2009-04-04 with grism 300V and filter GC
435, adopting a 0.4 $^{''}$ slit width. The blue solid lines in the left panels are the
measured $P_{Q}$ and $P_{U}$ values (in \%) obtained after correcting for
the respective instrumental polarization, and the combined wavelength
dependent effect for imperfections of the alignment of the fast axis of the retarder wave plate, and
 of the principal plane of the Wollaston prism. The solid lines in the right panels show the fraction of linear
polarization (upper panel), and the position
angle (bottom panel) corresponding to the $P_{Q}$ and $P_{U}$ values represented
in the left panels. According to \cite{Fossati07}, this
star has a fraction of linear polarization ({\rm wavelength dependent}) of 5.7\,\% in
the $B$ band, 6.3\,\% in the $V$ band, 6.1\,\% in the $R$ band, and a position
angle of about 179.5$^{\circ}$. Hence, for the measured values of the Stokes profiles, the fraction of linear polarization, and the
position angle of polarization of this star to be fully consistent with their respective values given by
\citet{Fossati07}, they have to be corrected for the FORS2 instrument offset
of the respective night (see Table \ref{table1}).

\begin{figure*}
  \centering
 \includegraphics[bb=65 550 550 705,clip,height=5.50cm,width=16.50cm]{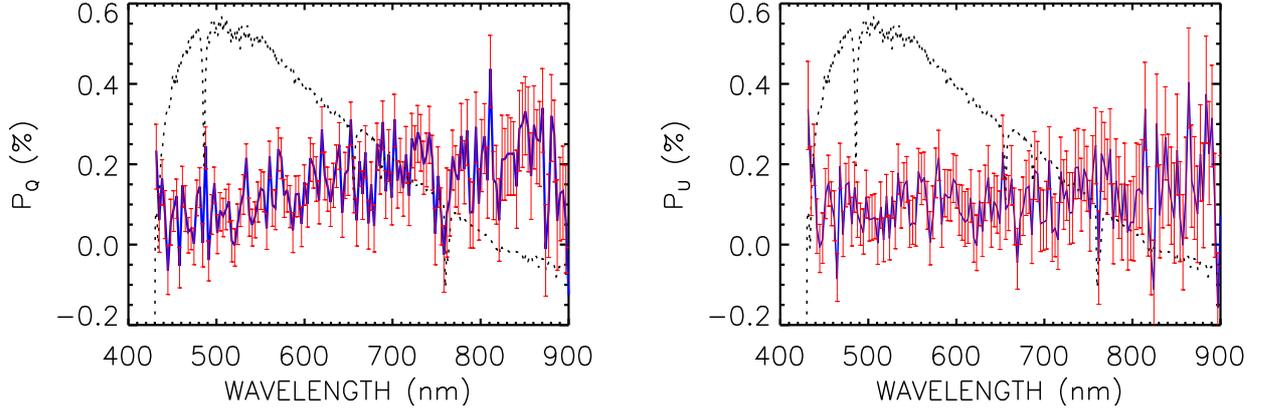}
 \caption{Spectro-polarimetry of the non-polarized standard star ${\rm HD\,97689}$
  observed with FORS2 during night 2010-05-04. The measured reduced Stokes $P_{Q}$ and $P_{U}$
  profiles, both with a value of $\sim 0.1$\,\% are indicated in solid blue lines, with the statistical error bars in red. In
 both panels, the black dotted lines show the total flux (in arbitrary
 units), and is not corrected for the instrument + telescope transmission functions.}
 \label{fig7}
  \end{figure*}

\begin{figure*}
  \centering
 \includegraphics[bb=65 370 550 705,clip,height=12.0cm,width=16.50cm]{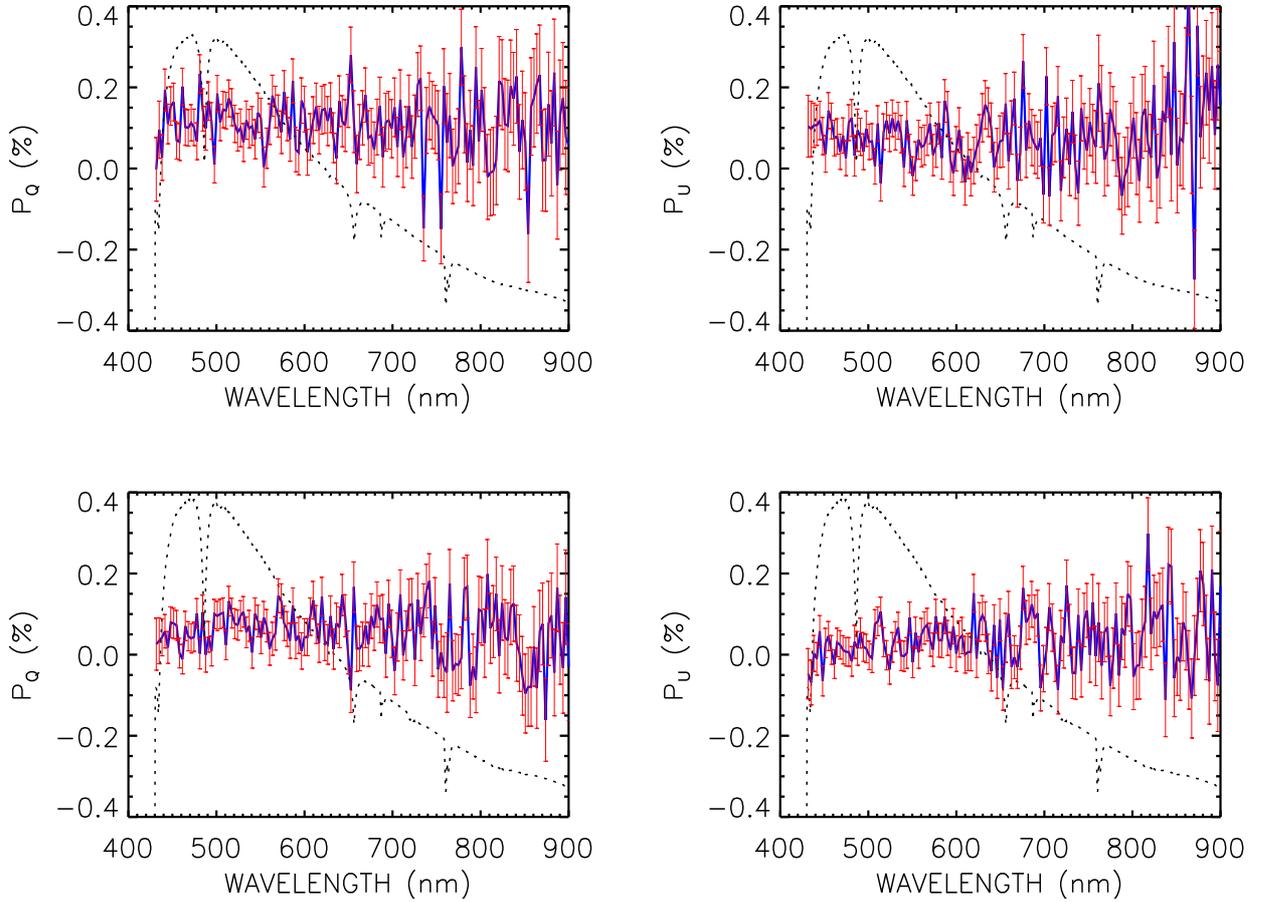}
 \caption{Same as Fig.~\ref{fig7}, for the non polarized standard star ${\rm WD\,1620391}$
  observed with FORS2 during nights 2010-07-25 (upper panel) and 2011-03-28 (lower panel). In the upper panels, it can be seen that the
  polarization value is $\sim$0.1\% both in Stokes $P_{Q}$ and $P_{U}$, while it is less in the case of lower panel, $\sim$0.05\%.}
 \label{fig8}
  \end{figure*}

\begin{figure*}
  \centering
 \includegraphics[bb=80 370 553 705,clip,height=12.00cm,width=16.50cm]{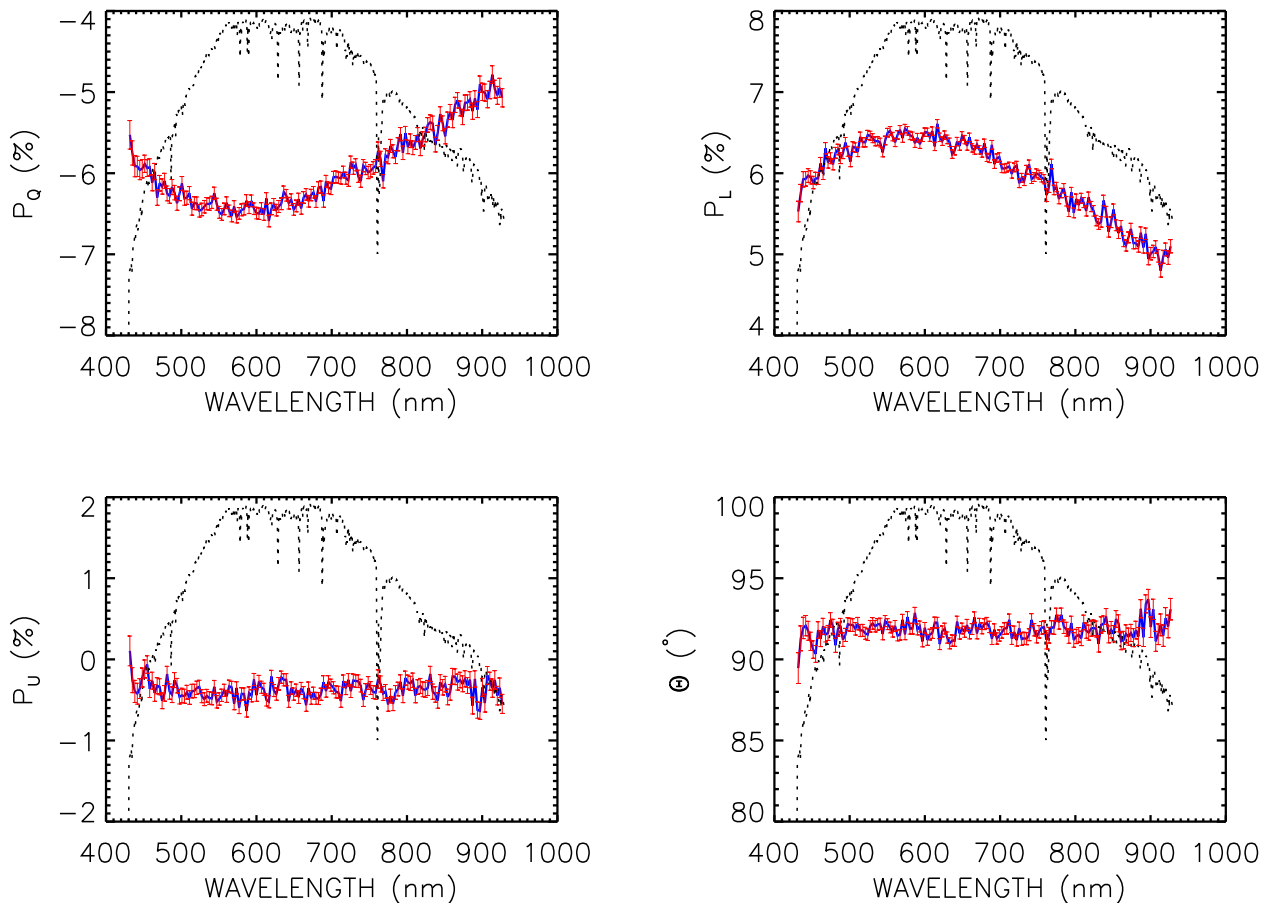}
 \caption{Spectro-polarimetry of the standard star for linear polarization
 CD-28 13479 (=Hiltner 652) observed during the night 2009-04-04. In the $P_{Q}$ and $P_{U}$
 panels, the blue solid lines represent the $P_{Q}$ and $P_{U}$ values after
 taking into account the combined correction due to
 the chromatism of the retarder waveplate and the Wollaston prism (i.e., wavelength dependent
 deviations of the position angle of the fast axis of the retarder waveplate and of the Wollaston
  prism from their respective nominal values), and the instrumental polarization; all discussed in the text.
  The error bars of each spectral point are represented in red. The $P_{\rm L}$ panel shows the total fraction of linear
 polarization, and the bottom right panel shows the position angle of the observed polarization (blue solid
 lines). These observed values have to be corrected for the FORS2 instrument offset of the respective epoch
 to be fully consistent with their expected values, measured with FORS1. All dotted black lines show Stokes $I$ expressed in arbitrary units.}
 \label{fig6}
  \end{figure*}
\begin{center}
\begin{table}
\caption{ ESO VLT FORS2 Instrument offset during different observation periods.}
\label{table1}
\begin{tabular}{ccccc}
\hline \hline
Epoch&Observed STD &$\Theta_{\rm measured}~(^{\circ})$ &$\Theta_{\rm real}~(^{\circ})$&$\Theta_{0}~(^{\circ})$\\
\hline
2009-04-04&Hiltner 652&91.9&179.5& -87.6\\
2010-02-22&Ve 6-23&158.2&172.5& -14.3\\
2010-05-04&Ve 6-23&175.0&172.5&+2.5\\
\hline
\end{tabular}
\end{table}
\end{center}
\subsection{Results for Iapetus}
Eq. (2) above conveys that the error bar for $P_{X}$ is equal to the inverse of the S/N of
 the flux accumulated in both beams ($f^{\parallel}$ + $f^{\perp}$) from all exposures obtained at all retarder wave plate
 positions. The signal-to-noise ratio we achieved during our observation varies from $\sim 830 - 1250 $~(on both beams at all positions of the retarder waveplate),
which is high enough to obtain polarization measurements with uncertainty
levels of $\sim 0.1$\% per spectral bin \citep[see e.g.,][]{Bagnulo06a,Bagnulo06b}.\\
As a validation of the reliability of our measurement uncertainities,we have also systematically calculated the null
 profiles, i.e., the difference between Stokes profiles obtained from different pairs of exposures, using the relation:
\begin{equation}
N_{X}=\frac{1}{2N} \sum_{j=1}^{N}(-1)^{(j-1)} \left [\left(  \frac{f^{\parallel}-f^{\perp}} {f^{\parallel}+f^{\perp}} \right )_{\alpha_{j}}   - \left ( \frac{f^{\parallel}-f^{\perp} } {f^{\parallel}+f^{\perp}} \right )_{\alpha_{j}+ \Delta P}\right],
\end{equation}
where $\Delta$P is the difference in the position of the retarder waveplate
between two consecutive exposures, and is equal to 45$^{\circ}$ for the linear polarization
measurements, and 90$^{\circ}$ in the case of circular
polarization measurements. These expressions are valid only for an even
number of pairs of exposures at different position angles, and could be
calculated only for the scientific observations of Iapetus, but not for the
polarimetric standard stars discussed above, that were observed only at two position angles of the
retarder wave plate for each Stokes parameter.
The null profiles should have a Gaussian distribution centered about $0$ with
the same $\sigma$ given by Eq. (3). For further technical details of null parameters, we refer the reader to
  \citet{Bagnulo09}.
\\Adopting the perpendicular
to the great circle passing through Iapetus and the Sun as the reference
direction, as it was the case in \citet{Bagnulo06a} and \citet{LandiDegl'lnnocenti07}, our measurements are transformed from the instrument reference system (the north celestial pole) to the reference system with the $x$-axis (the reference direction) perpendicular to the scattering
plane, according to the relation:
\begin{equation}
 \begin{aligned}
P_{Q}^{'} &=(P_{Q}-q_{\rm ins})\cos(2\Theta)+  P_{U}\sin(2\Theta),\\
P_{U}^{'} &=-(P_{Q}-q_{\rm ins})\sin(2\Theta)+  P_{U}\cos(2\Theta),\\
P_{V}^{'}&=P_{V}
\end{aligned}
 \end{equation}
\noindent with $\Theta =\phi+\frac{\pi }{2}+\Delta +\Theta_{0}$, where $\Theta_{0}$ is the FORS2 instrument offset angle during each
 epoch of our observation as indicated in Table~\ref{table1}, and $\phi$ is the angle between the celestial meridian passing through
Iapetus and the great circle passing through the Sun and Iapetus (increased by $90^{\circ}$, so as to have
 the reference direction perpendicular to the scattering plane).
Using Eq. (10) of \citet{LandiDegl'lnnocenti07}, for the nights 2009-04-04, 2010-02-22, 2010-05-04, and 2010-07-25 of our observations, we calculated  $\phi$ values (at the middle of the series of exposures) of
$-62.95^{\circ}$, $-70.94^{\circ}$,$-64.08^{\circ}$, and $-67.92^{\circ}$,
respectively. The quantity $\Delta$ is a combined value for deviations of the fast axis of the retarder waveplate
and of the principal plane of the Wollaston prism from their nominal values (both wavelength
dependent, \citep[see e.g.,][]{Bagnulo09}, and is tabulated
in the ESO instrument webpage \footnote[1]{ (http://www.eso.org/sci/facilities/paranal/instruments/fors/inst/pola.html)}.
The quantity $q_{\rm ins}$ is a constant polarization value we set to $\sim$0.1\% (as discussed above), and accounts for an instrumental contribution to the
observed polarization in $P_{Q}$ (in the instrument reference system) that seems to affect
spectro-polarimetric observations \citep{Fossati07}. Without adding it to
$P_{Q}$, we would measure an offset in the $P_{U}$ profile from zero by a (small)
constant value, which is in fact consistent with what was observed with FORS1 by
\citet{Fossati07}. In Eq. (4) above, the prime index (~${}'$~) is used to refer to the profiles measured in the reference system associated to the scattering
plane.

Figure~\ref{fig1} shows linear spectro-polarimetric measurement of Iapetus' bright side at a single observing epoch.
In the left panels,  The red lines in the left panels represent the measured $P_{Q}^{'}$ and $P_{U}^{'}$ values (in \%) after applying
 the transformation relation given in Eq. (4). The respective null profiles,
 $N_{U}$ (offseted to $+ 0.25$\,\% for display purposes), and $N_{Q}$
 are are shown as blue solid lines, superposed to the statistical error bars
 reporeted in light blue. This
allows one to immediately visualize the points of the null profiles that are different from zero
by a quantity higher than the error bar, and thus, to evaluate the reliability of the corresponding points
of the Stokes profiles. The right panels of Fig.~\ref{fig1}, show the total fraction of
linear polarization (top panel), and the observed position angle. As clearly demonstrated in Fig.~\ref{fig1}, when $P_{U}^{'}$
 equals zero, the position angle of the polarization plane, computed using Eq. (6) of \citet{Bagnulo06a}, and measured from the axis perpendicular to the scattering plane, equals $90^{\circ}$.\\
To compare our linear spectro-polarimetric measurements of Iapetus with the imaging polarimetry of other
objects, we convolved the obtained polarized spectra with the transmission function of the $BVRI$ Bessel filters  applying the
convolution relation given by Eq. (12) of \citet{Fossati07}, and the results are depicted in
Fig.~\ref{fig5}.\\
Figure \ref{fig9new} shows measurement of our circular polarization at a single phase angle.
The red solid line represents the average $P_{V}^{'}$ profile (in \%) obtained from instrument position
angles that differ by $90^{\circ}$. The blue dashed line  represents measurement at $0 ^{\circ}$
instrument position angle, while the dotted green line is the measurement at $90^{\circ}$ instrument position
angle on the sky. The average null profile is found to be scattered around zero within the corresponding error bars of
the average Stokes $P_{V}^{'}$ profile, thereby confirming the reliability of our measurement.

\section{Discussions}
\begin{figure*}
  \centering
 \includegraphics[bb=65 370 553 718,clip,height=10.0cm,width=13.85cm]{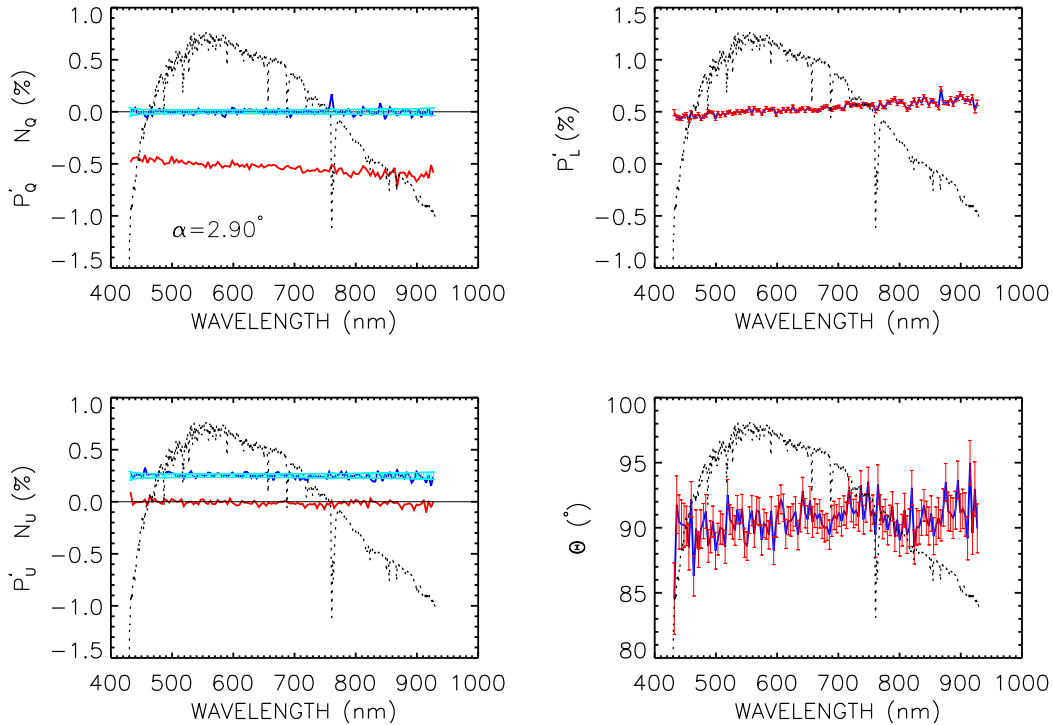}
 \caption{Linear polarization of the bright side of Iapetus versus wavelength, observed with FORS2
 at the ESO VLT in April 2009 at phase angle $2.90^{\circ}$.
In the $P_{Q}^{'}$ and $P_{U}^{'}$ panels, the red lines represent the measurements
corrected for the combined offset due to the chromatism of the retarder waveplate + the Wollaston prism, the instrumental
polarization, the FORS2 instrument offset of the respective epoch, and
transformed in to a system where the reference direction is the direction
perpendicular to the plane identified by the Sun, the object, and the
observer (the scattering plane). The null profiles, $N_{U}$ (offseted to $+ 0.25$\,\% for display purpuses) and
$N_{Q}$ are displayed in blue, superposed to the statistical error bars (in light blue). Ideally, null
 profiles should be centered about zero with a Gaussian distribution that has
 the same $\sigma$ as the corresponding Stokes parameters. Therefore, any point of the null
 profiles that significantly exceeds the light blue error bar suggests that
 the spectral points in the corresponding Stokes profiles are unreliable. The
  right panels show the fraction of linear polarization and its position angle, together with their error bars. For
  symmetric reasons, we expect the position angle to be centred about $90^{\circ}$ (as well as we expect $P_{U}^{'}$ to be centred about zero). In
 all panels, the black dotted lines show the total flux (in arbitrary units).}

\label{fig1}
  \end{figure*}
\begin{figure*}
\centering
\includegraphics[bb=61 549 301 705,clip,height=5.50cm,width=8.0cm]{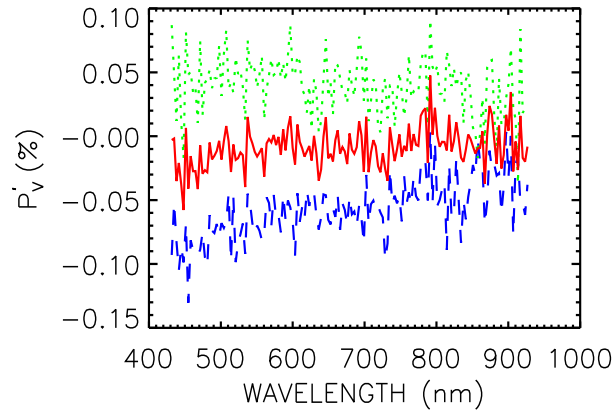}
\caption{Circular polarization of the bright side of Iapetus versus wavelength, obtained with
FORS2 at the ESO VLT during the night 2011-03-28 at a phase angle of $0.77^{\circ}$. The
blue dashed curve shows the Stokes $P_{V}^{'}$ profile obtained at $0.0^{\circ}$ instrument position angle (with
respect to the north celestial pole), the green dotted curve showing the measurement obtained after rotating the instrument by
$90.0^{\circ}$ on the sky, and the red solid curve represents the average of the two signals. It can
be noted that a change in signal occurs in circular polarization observations that are taken after rotating the instrument by 90
degree with respect to the intial position angle on the sky.}
\label{fig9new}
\end{figure*}
\begin{figure*}
  \centering
  \includegraphics[bb=65 370 553 718,clip,height=10.0cm,width=13.85cm]{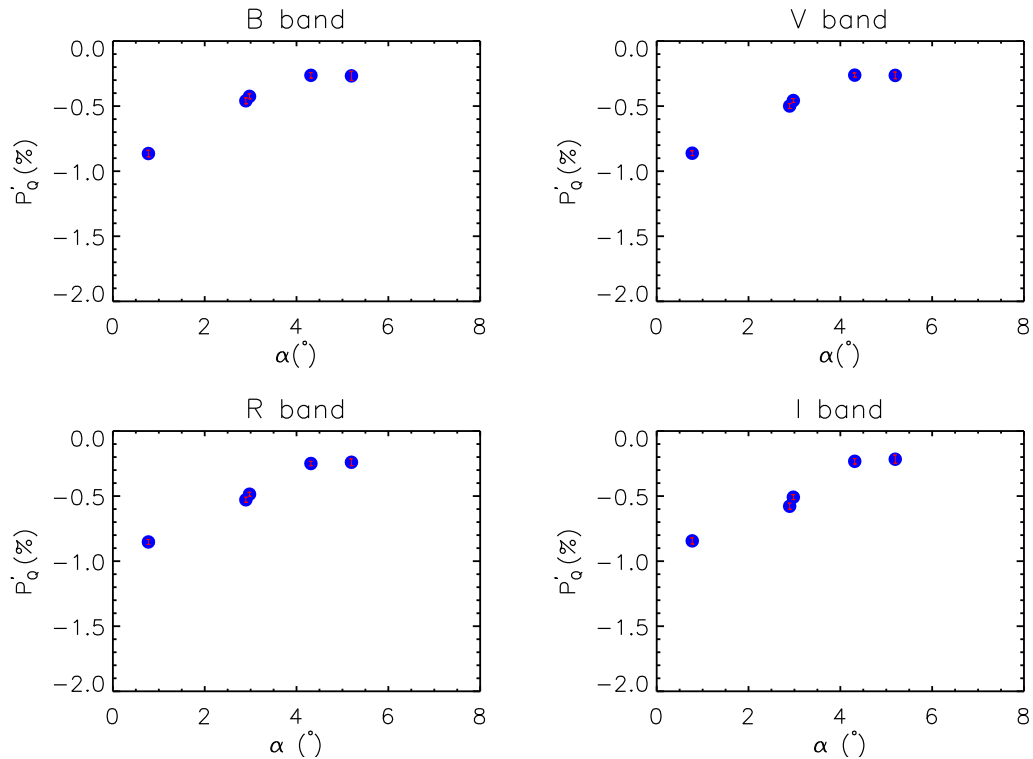}
 \caption{$P_{Q}^{'}$ values for Iapetus (obtained after convolving the polarized spectra) versus the phase angle of observation in four Bessel filter bands.}
 \label{fig5}
   \end{figure*}

\begin{center}
\begin{table*}
\caption{Results of polarimetric observations of the bright side of Iapetus,
containing the epoch of the observations (date and UT time), the exposure time, the phase angle,
the measured Stokes parameters $P_{Q}^{'}$ and $P_{U}^{'}$ (with respect to the scattering plane), and the longitude of the
 sub-Earth point.}
\begin{tabular}{cccccccc}
  \hline \hline
Date\\(yyyy mm dd) &Time (UT) (hh:mm)&Exp (sec)&$\alpha~(^{\circ}$)&Filter&$P_{Q}^{'}$(\%)&$P_{U}^{'}$(\%)& Ob-lon ($^{\circ}$)\\
\hline
    2009-04-04&05:04 - 05:19&$20\times16$& 2.90&$B$& $-0.46\pm0.02$&$0.00\pm0.02$& 271.36\\
                       &                        &      &       & $V$& $-0.50\pm0.02$ &$ 0.00\pm0.02$    \\
                       &                       &       &           &$R$& $-0.53\pm0.02$       &$-0.01\pm0.02$\\
                         &                   &          &         &$I$&$-0.58\pm0.02$        & $ -0.02\pm0.02$   \\

 2010-02-22&04:11 - 04:29 &$40\times16$& 3.0&$B$&$ -0.42\pm 0.02$&$0.00\pm0.02$&284.86\\
                      &                        &       &    &$V$&$ -0.46\pm 0.02$ &$ -0.02\pm0.02$    \\
                       &                        &      &   &$R$&$ -0.49\pm 0.02$       &$ -0.03\pm0.02$    \\
                         &                   &         &  &$I$&$ -0.51 \pm 0.02$        &$-0.02\pm0.02$     \\

2010-05-04&00:02 - 00:51&$50\times16$& 4.31&$B$&$ -0.26\pm 0.02$&$0.04\pm0.02$&251.13\\
                    &                      &        &     &$V$&$ -0.26\pm 0.02$      & $0.02\pm0.02$    \\
                       &                   &         &     &$R$&$ -0.25\pm 0.02$       &$-0.01\pm0.02$  \\
                       &                   &          &     &$I$&$  -0.23\pm 0.02$        &$-0.04\pm0.02$     \\

2010-07-25&23:55 - 00:37  &$50\times16$& 5.20&$B$&$ -0.27\pm 0.03$&$0.00\pm0.03$&261.40\\
                        &                    &        &      &$V$&$  -0.26\pm 0.03$  &$ -0.02\pm0.03$    \\
                        &                   &          &      &$R$&$  -0.24\pm 0.03$      &$ -0.04\pm0.03$    \\
                        &                   &          &       &$I$& $ -0.22\pm 0.03$        &$ -0.05\pm0.03$    \\

2011-03-28&08:07 - 09:33 &$20\times 16$& 0.77&$B$&$-0.86\pm 0.03$&$0.00\pm 0.03$&285.26\\
          &              &             &      &$V$&$  -0.86\pm 0.02$ &$0.00\pm 0.02$\\
          &              &             &      &$R$&$  -0.85\pm 0.02$ &$0.00\pm 0.02$\\
          &              &             &      &$I$&$  -0.84 \pm 0.02$&$0.00\pm 0.02$\\
\hline
\end{tabular}
\footnotetext{Measurements in different filter bands were obtained after applying eq.(6).}
\label{table2}
\end{table*}
\end{center}
Our linear spectro-polarimetric measurements of Iapetus' bright side against
their respective phase angles of observations are given in Fig.~\ref{fig5}. The measurements show that the linear
degree of Iapetus' bright side polarization changes with increasing phase angle,
  from $\sim$$-0.90$\,\% at $0.77^{\circ}$ to $\sim$$-0.30$\,\% at $5.20^{\circ}$.
The polarization of Iapetus' bright side at $1.0^{\circ}$ phase angle
obtained by \citet{Rosenbush02} was $\sim$ $-0.8$\,\%, and the value measured at phase angles between 5.0 and $6.0^{\circ}$
was $\sim$$-0.2$\,\%  \citep[see][]{Zellner72,Rosenbush02,Veverka77}.
 The later polarization value is believed to be typical for E-type asteroids 44 Nysa \citep{Rosenbush09} and 64 Angelina \citep{Rosenbush05b}, and also for
the Galilean satellites, Europa and Ganymede \citep{Rosenbush1997}, at similar phase angles. Thus, our
measurements (at similar phase angles), confirm the same notion as depicted in Table~\ref{table2}.
 It is worth noting, in particular, that the largest linear polarization at all
wavelengths we obtained is at the  smallest phase angle $0.77^{\circ}$. The
coherent backscattering model \citep{Mishchenko2000} suggests
that polarizations as large as -2.8\% can occur near $1.0^{\circ}$ phase
angle, and such trends of dramatic change in polarization  have been observed for some of the Galilean satellites at phase angles of about
$0.6-0.7^{\circ}$ \citep[see][]{Rosenbush1997}. Hence, our measurement at the $0.77^{\circ}$ phase angle is well consistent with such polarization trend.\\
As for the wavelength dependence of polarization, the variation of Iapetus' bright side polarization with wavelength
is in general less pronounced (see Table~\ref{table2}). \citet{Rosenbush1997} have also
reported that the spectral dependence of polarization of Io, Europa and Ganymede
in the $BVR$ filters is weak, and hence, our measurements confirm that such
a small wavelength dependence of polarization might be typical for some satellites.
However, it is not obvious at present to judge if it offers any significant constraints on the nature of the scattering surface, although
\citet{Belskaya09} have pointed out that, in the case of asteroids, the variation in the observed wavelength dependence of polarization
seems to be attributed mainly to the surface composition.\\
 It has been revealed that the larger TNOs like Pluto and Eris display shallow polarimetric phase functions
, almost constant over the observed phase angle range \citep{Bagnulo08}, and possibly indicative for surfaces dominated by
   large (compared to wavelength) inhomogeneous particles \citep{Belskaya08}, while the smaller ones show rather
  steep negative polarization phase functions most likely produced by a mixture of at least two
  different types of smaller (10-100\,$\mu$m size) grains of  single scattering albedo \citep{Boehnhardt04,
  Bagnulo06a}.\\Comparison of our measurements with those of TNOs
\citep[]{Boehnhardt04,Bagnulo06a,Bagnulo08,Belskaya08}, {\textit{within the same phase-angle range}, has shown that the polarization of Iapetus'
bright hemisphere seems to be slightly deeper than that of the larger TNOs, and it tends to have a similar value as that of smaller ones particularly
around a phase angle of $\sim$ 1.0$^{\circ}$.
 Larger TNOs are believed to have
surfaces dominated by either methane ice or water ice \citep[see, e.g.,][]{Barucci08, Brown08}, and hence,
the fact that Iapetus' polarization is deeper than that of these objects clearly demonstrates that polarization is strongly sensitive to a variety of the
scattering surface parameters such as particle size, composition, shape (structure), surface texture, refractive indices, etc.,
apart from surface reflectance of objects, and it is almost
impossible to interpret such differences uniquely.
Compared to polarimetric measurements Centaurs \citep[]{Bagnulo06a, Belskaya10}, Iapetus' polarization is shallower than that
 of any of the three Centaurs (Chiron, Pholus, and Chariklo), in agreement with the fact that
 darker surfaces exhibit deeper linear degree of negative polarization than the brighter ones, since these objects are
 of lower albedo than the trailing hemisphere of Iapetus. \\ Comparison of our measurements with that of the nucleus of
 comet 2P/Encke \citep{Boehnhardt:08} and the main belt object 133P/Elst-Pizarro \citep{Bagnulo:10} around similar phase angle range of observations,
  shows that Iapetus has a completely different polarimetric behaviour since the
  polarimetric phase function of the nucleus of
 comet 2P/Encke was found to be different from any of other
  solar system objects, while that of 133P/Elst-Pizarro resembles more likely that of
  F-type asteroids than icy bodies.\\Our measurements of circular polarization of Iapetus' bright side (see Fig. \ref{fig9new}) had the goal to test
 whether optically active, possibly organic surface material is present on the bright hemisphere that is believed to induce
 circular polarization of a few percent due to its chiralic character (Sterzik et al. 2010, Rosenbush et al. 2007).
Our observed circular polarization depends on the instrument position angle, and in particular it changes
in sign when the instrument is rotated by $90^{\circ}$(see Fig. \ref{fig9new}). As discussed by
 \citet{Bagnulo09}, this is attributed to an instrumental effect, and most likely it is a symptom of cross-talk from linear to circular polarization.
 If we assume that the only instrumental effect is pure cross-talk from linear to circular polarization, then the average of the $P_{V}^{'}$ profiles
  observed at the two instrument position angles (solid red line in Fig.~5) {\it should} be intrinsic to the
  source. In a strict sense, this value is a non-zero, as in particular can be seen clearly around shorter wavelengths.
  Hence, given the fact that circular polarization is typically small \citep[see, e.g.,][]{Rosenbush97b}, such a small non-zero value of our circular polarization
   is not unexpected. However, we must also remark that the FORS instrument lacks a sufficiently accurate
      characterisation to validate a detection of circular polarization at the level of a few units in $10^{-4}$ in
       the continuum, even using our specific observing technique.
\section{Conclusions}
We have measured the spectral polarization of the bright side
of Iapetus with an accuracy better than $\sim 0.1$\,\%. The linear degree of negative polarization decreases with increasing phase angle
from $-0.9$\,\% at $0.77^{\circ}$ to $-0.3$\,\% at $5.2^{\circ}$. The magnitude of the polarization increases monotonically with
wavelength across the wavelength range. The fact that our polarimetric phase function of Iapetus is fully
in agreement with previous measurements \citep[]{Zellner72,Rosenbush02,Veverka77}, and with other high albedo objects (icy Galilean satellites, E-type
asteroids, etc.), it offers an additional line of evidence for the light scattering behaviour of small solar system bodies that
have high surface albedo and (or) that have surfaces rich in water ice. To confirm the notion of \citet{Rosenbush02} that Iapetus' bright hemisphere
displays two branches of negative polarization, one called the polarization opposition
effect (occurring at very small phase angles), and the other one called the normal
branch of negative polarization (at relatively higher phase angles), measurements at low phase angles would be highly
needed. Apart from being a test for optically active (chiral) organic molecules, circular
polarization is also believed to be sensitive to the shape, structure and
composition of the scattering surfaces \citep{Rosenbush07}, and thus, our
measurement of circular polarization of Iapetus' bright hemisphere
combined with its linear polarization measurements would allow one to translate this polarimetric characteristic of typical water ice to the light scattering
behaviour of other solar system bodies of dominant water ice surface
constitution.
\begin{acknowledgements}
The authors sincerely thank Dr. Jay Goguen (the referee) for the help in his useful and detailed report to significantly
improve the quality of this paper. C. Ejeta acknowledges the PhD studentiship supported by the Helmholtz Association through
the research alliance 'Planetary Evolution and Life', and he thanks Armagh observatory for the kind hospitality during his research visits.
\end{acknowledgements}
\bibliographystyle{aa}
\bibliography{15491_bibtex}

\begin{thebibliography}{27}
\expandafter\ifx\csname natexlab\endcsname\relax\def\natexlab#1{#1}\fi

\bibitem[{{Appenzeller} {et~al.}(1998){Appenzeller}, {Fricke}, {F{\"u}rtig},
  {G{\"a}ssler}, {H{\"a}fner}, {Harke}, {Hess}, {Hummel}, {J{\"u}rgens},
  {Kudritzki}, {Mantel}, {Meisl}, {Muschielok}, {Nicklas}, {Rupprecht},
  {Seifert}, {Stahl}, {Szeifert}, \& {Tarantik}}]{Appenzeller98}
{Appenzeller}, I., {Fricke}, K., {F{\"u}rtig}, W., {et~al.} 1998, The
  Messenger, 94, 1

\bibitem[{{Bagnulo} {et~al.}(2008){Bagnulo}, {Belskaya}, {Muinonen}, {Tozzi},
  {Barucci}, {Kolokolova}, \& {Fornasier}}]{Bagnulo08}
{Bagnulo}, S., {Belskaya}, I., {Muinonen}, K., {et~al.} 2008, \aap, 491, L33

\bibitem[{{Bagnulo} {et~al.}(2006{\natexlab{a}}){Bagnulo}, {Boehnhardt},
  {Muinonen}, {Kolokolova}, {Belskaya}, \& {Barucci}}]{Bagnulo06a}
{Bagnulo}, S., {Boehnhardt}, H., {Muinonen}, K., {et~al.} 2006{\natexlab{a}},
  \aap, 450, 1239

\bibitem[{{Bagnulo} {et~al.}(2009){Bagnulo}, {Landolfi}, {Landstreet}, {Landi
  Degl'lnnocenti}, {Fossati}, \& {Sterzik}}]{Bagnulo09}
{Bagnulo}, S., {Landolfi}, M., {Landstreet}, J.~D., {et~al.} 2009, \pasp, 121,
  993

\bibitem[{{Bagnulo} {et~al.}(2006{\natexlab{b}}){Bagnulo}, {Landstreet},
  {Mason}, {Andretta}, {Silaj}, \& {Wade}}]{Bagnulo06b}
{Bagnulo}, S., {Landstreet}, J.~D., {Mason}, E., {et~al.} 2006{\natexlab{b}},
  \aap, 450, 777

\bibitem[{{Bagnulo} {et~al.}(2010){Bagnulo}, {Tozzi}, {Boehnhardt}, {Vincent},
  \& {Muinonen}}]{Bagnulo:10}
{Bagnulo}, S., {Tozzi}, G.~P., {Boehnhardt}, H., {Vincent}, J.-B., \&
  {Muinonen}, K. 2010, \aap, 514, A99

\bibitem[{{Barucci} {et~al.}(2008){Barucci}, {Brown}, {Emery}, \&
  {Merlin}}]{Barucci08}
{Barucci}, M.~A., {Brown}, M.~E., {Emery}, J.~P., \& {Merlin}, F. 2008,
  {Composition and Surface Properties of Transneptunian Objects and Centaurs},
  ed. {Barucci, M.~A., Boehnhardt, H., Cruikshank, D.~P., Morbidelli, A., \&
  Dotson, R.}, 143--160

\bibitem[{{Belskaya} {et~al.}(2008){Belskaya}, {Bagnulo}, {Muinonen},
  {Barucci}, {Tozzi}, {Fornasier}, \& {Kolokolova}}]{Belskaya08}
{Belskaya}, I., {Bagnulo}, S., {Muinonen}, K., {et~al.} 2008, \aap, 479, 265

\bibitem[{{Belskaya} {et~al.}(2010){Belskaya}, {Bagnulo}, {Barucci},
  {Muinonen}, {Tozzi}, {Fornasier}, \& {Kolokolova}}]{Belskaya10}
{Belskaya}, I.~N., {Bagnulo}, S., {Barucci}, M.~A., {et~al.} 2010, \icarus,
  210, 472

\bibitem[{{Belskaya} {et~al.}(2009){Belskaya}, {Levasseur-Regourd}, {Cellino},
  {Efimov}, {Shakhovskoy}, {Hadamcik}, \& {Bendjoya}}]{Belskaya09}
{Belskaya}, I.~N., {Levasseur-Regourd}, A.-C., {Cellino}, A., {et~al.} 2009,
  \icarus, 199, 97

\bibitem[{{Boehnhardt} {et~al.}(2004){Boehnhardt}, {Bagnulo}, {Muinonen},
  {Barucci}, {Kolokolova}, {Dotto}, \& {Tozzi}}]{Boehnhardt04}
{Boehnhardt}, H., {Bagnulo}, S., {Muinonen}, K., {et~al.} 2004, \aap, 415, L21

\bibitem[{{Boehnhardt} {et~al.}(2008){Boehnhardt}, {Tozzi}, {Bagnulo},
  {Muinonen}, {Nathues}, \& {Kolokolova}}]{Boehnhardt:08}
{Boehnhardt}, H., {Tozzi}, G.~P., {Bagnulo}, S., {et~al.} 2008, \aap, 489, 1337

\bibitem[{{Brown}(2008)}]{Brown08}
{Brown}, M.~E. 2008, {The Largest Kuiper Belt Objects}, ed. {Barucci, M.~A.,
  Boehnhardt, H., Cruikshank, D.~P., Morbidelli, A., \& Dotson, R.}, 335--344

\bibitem[{{Fossati} {et~al.}(2007){Fossati}, {Bagnulo}, {Mason}, \& {Landi
  Degl'lnnocenti}}]{Fossati07}
{Fossati}, L., {Bagnulo}, S., {Mason}, E., \& {Landi Degl'lnnocenti}, E. 2007,
  in Astronomical Society of the Pacific Conference Series, Vol. 364, The
  Future of Photometric, Spectrophotometric and Polarimetric Standardization,
  ed. {C.~Sterken}, 503

\bibitem[{{Izzo} {et~al.}(2010){Izzo}, {de Bilbao}, {Larsen}, {Bagnulo},
  {Freudling}, {Moehler}, \& {Ballester}}]{Izzo10}
{Izzo}, C., {de Bilbao}, L., {Larsen}, J., {et~al.} 2010, in Society of
  Photo-Optical Instrumentation Engineers (SPIE) Conference Series, Vol. 7737,
  Society of Photo-Optical Instrumentation Engineers (SPIE) Conference Series

\bibitem[{{Landi Degl'lnnocenti} {et~al.}(2007){Landi Degl'lnnocenti},
  {Bagnulo}, \& {Fossati}}]{LandiDegl'lnnocenti07}
{Landi Degl'lnnocenti}, E., {Bagnulo}, S., \& {Fossati}, L. 2007, in
  Astronomical Society of the Pacific Conference Series, Vol. 364, The Future
  of Photometric, Spectrophotometric and Polarimetric Standardization, ed.
  {C.~Sterken}, 495

\bibitem[{{Mishchenko} {et~al.}(2000){Mishchenko}, {Luck}, \&
  {Nieuwenhuizen}}]{Mishchenko2000}
{Mishchenko}, M., {Luck}, J.-M., \& {Nieuwenhuizen}, T. 2000, Journal of
  optical society of America.A, 17, 888

\bibitem[{{Rosenbush} {et~al.}(2002){Rosenbush}, {Kiselev}, {Avramchuk}, \&
  {Mishchenko}}]{Rosenbush02}
{Rosenbush}, V., {Kiselev}, N., {Avramchuk}, V., \& {Mishchenko}, M. 2002, in
  Optics of Cosmic Dust, ed. {G.~Videen \& M.~Kocifaj}, 191

\bibitem[{{Rosenbush} {et~al.}(2007){Rosenbush}, {Kolokolova}, {Lazarian},
  {Shakhovskoy}, \& {Kiselev}}]{Rosenbush07}
{Rosenbush}, V., {Kolokolova}, L., {Lazarian}, A., {Shakhovskoy}, N., \&
  {Kiselev}, N. 2007, \icarus, 186, 317

\bibitem[{{Rosenbush} {et~al.}(1997{\natexlab{a}}){Rosenbush}, {Avramchuk},
  {Rosenbush}, \& {Mishchenko}}]{Rosenbush1997}
{Rosenbush}, V.~K., {Avramchuk}, V.~V., {Rosenbush}, A.~E., \& {Mishchenko},
  M.~I. 1997{\natexlab{a}}, \apj, 487, 402

\bibitem[{{Rosenbush} {et~al.}(2005){Rosenbush}, {Kiselev}, {Shevchenko},
  {Jockers}, {Shakhovskoy}, \& {Efimov}}]{Rosenbush05b}
{Rosenbush}, V.~K., {Kiselev}, N.~N., {Shevchenko}, V.~G., {et~al.} 2005,
  \icarus, 178, 222

\bibitem[{{Rosenbush} {et~al.}(1997{\natexlab{b}}){Rosenbush}, {Shakhovskoj},
  \& {Rosenbush}}]{Rosenbush97b}
{Rosenbush}, V.~K., {Shakhovskoj}, N.~M., \& {Rosenbush}, A.~E.
  1997{\natexlab{b}}, Earth Moon and Planets, 78, 381

\bibitem[{{Rosenbush} {et~al.}(2009){Rosenbush}, {Shevchenko}, {Kiselev},
  {Sergeev}, {Shakhovskoy}, {Velichko}, {Kolesnikov}, \&
  {Karpov}}]{Rosenbush09}
{Rosenbush}, V.~K., {Shevchenko}, V.~G., {Kiselev}, N.~N., {et~al.} 2009,
  \icarus, 201, 655

\bibitem[{{Spencer} \& {Denk}(2010)}]{Spencer10}
{Spencer}, J.~R. \& {Denk}, T. 2010, Science, 327, 432

\bibitem[{{Squyres} \& {Sagan}(1983)}]{Squyres:1983}
{Squyres}, S.~W. \& {Sagan}, C. 1983, \nat, 303, 782

\bibitem[{{Veverka}(1977)}]{Veverka77}
{Veverka}, J. 1977, in IAU Colloq. 28: Planetary Satellites, ed. {J.~A.~Burns},
  210--230

\bibitem[{{Zellner}(1972)}]{Zellner72}
{Zellner}, B. 1972, \apjl, 174, L107

\end{thebibliography}
\end{document}